\documentclass[
  aps,
  prd,
  10pt,
  twocolumn,
  superscriptaddress,
  nofootinbib,
  amsmath,
  amssymb
]{revtex4-2}

\usepackage{graphicx}
\usepackage{eqnarray}
\usepackage{amsfonts,amssymb}
\usepackage[hidelinks]{hyperref}

\begin{document}

\title{Matter density perturbations in quadratic gravity}

\author{Pedro Bessa}
\email{pedvbessa@gmail.com}
\affiliation{Centro Brasileiro de Pesquisas F\'isicas, Rua Dr. Xavier Sigaud,
150, Urca, Rio de Janeiro, RJ, Brazil}

\date{\today}

\begin{abstract}
We investigate linear matter-density perturbations in full quadratic gravity. Working in the
longitudinal gauge and adopting the sub-horizon and quasi-static
approximations, we obtain an algebraic expression for the effective gravitational coupling in the theory
and the equation for evolution of matter density perturbations. We identify the positive-real poles of this coupling
and map their dependence on the quadratic couplings. In the formal infrared
continuation, the reduced coupling recovers its general-relativistic value,
whereas at large physical wavenumber it is suppressed when both quadratic
couplings are nonzero; the single-coupling limits reproduce distinct
behaviors. Numerical integrations of the growing mode are performed up to
the first pole and show that pole-free regions of the reduced system can
support regular matter-growth solutions, while poles obstruct a continuous
late-time evolution. Our results show that quadratic gravity cannot generally reproduce the standard
cosmological evolution of perturbations and in the regions where it can, it closely matches the behavior of
$f(R)$ theories.
\end{abstract}

\maketitle
\newcommand{\revision}[1]{#1}

\section{Introduction}
In the past couple of decades, with the discovery of the accelerated expansion of the Universe, the field of \revision{modified gravity} has been extensively studied as a possible alternative to the $\Lambda$CDM standard model of cosmology, and manifold models beyond \revision{general relativity} have been proposed as possible theories to describe gravity at the largest scales without assuming \revision{dark energy} or exotic energy components in Einstein's equations \cite{Clifton:2011jh,CANTATA:2021asi}. 

More recently, as of today unsolved tensions in cosmology such as the Hubble tension \cite{DiValentino:2021izs}, and observational signatures pointing to a dynamical dark sector \cite{DESI:2025fii} have further encouraged the pursuit of new and old gravitational theories in order to accommodate these different observations and alleviate the tensions by going beyond $\Lambda$CDM cosmology \cite{Odintsov:2020qzd,Plaza:2025gcv,Chudaykin:2024gol,DiValentino:2025otz}.

In this context, theories which include curvature scalars of higher order than $R$ in the action, \revision{such as $f(R)$ theories and Weyl gravity}, have been extensively explored as alternatives to the $\Lambda$CDM paradigm as well as alternatives to early-time dynamics and the standard inflationary scenario \cite{Wang:2020dsc,DeFelice:2010aj,Harko:2024fnt}. One such theory, which includes all terms quadratic in curvature invariants in the action, is quadratic gravity. The renormalizability of \revision{quadratic} gravity, first proved by Stelle \cite{PhysRevD.16.953}, made it a potential theory for quantum gravity and an ultraviolet completion of general relativity, although theoretical difficulties arise at both the classical and quantum levels when developing solutions for the theory \cite{Salvio:2018crh}.

By including terms quadratic in curvature, the equations of motion for quadratic gravity are fourth order in the field, and Ostrogradsky's theorem implies that the theory \revision{possesses} intrinsic instabilities or, equivalently, ghost modes of propagation \cite{Salvio:2018crh,Woodard:2015zca}. Such instabilities render it difficult to obtain general properties of the theory, since stable solutions may only be found in certain parts of the theory's parameter space and for certain solutions. In the case of cosmology, the fact that the theory is equivalent to a scalar--tensor theory for the FLRW metric motivates its study as \revision{a} gravitational theory describing the large-scale evolution of the Universe \cite{Schmidt:2006jt}, although this equivalence is valid only at the background level.

Cosmological scenarios in quadratic gravity (QG) in the early Universe have been extensively studied, where the theory is understood as a high-energy completion of general relativity (GR) and Starobinsky inflation \cite{Barrow:2006xb,Barrow:2007pm,Schmidt:2006jt}. In particular, the analysis of how such theories lead to stable de Sitter solutions in the early Universe in order to produce a generic inflationary phase \cite{Schmidt:2006jt} and a fixed de Sitter attractor \cite{Barrow:2007pm} have been previously explored. The stability of perturbative cosmological solutions has also been studied at early cosmological times \cite{Asorey:2024oxw,Barrow:2007pm}, and more recently there has been extensive work on how proper gauges can lead to \revision{a} better understanding of the theory's extra degrees of freedom \cite{Alves:2025zkk,Palomares:2026iwt}. Literature that deals with the behavior of the theory at late times, where pressureless matter dominates the energy content of the Universe, is still in development, and works such as \cite{Asorey:2024oxw} provide results in the direction of the asymptotic behavior of the theory at late times.

In this work we obtain the evolution equation for matter perturbations in QG, derive an analytical expression for the \revision{effective gravitational coupling $G_{\rm eff}$}, and analyze the stability of such perturbations both analytically and numerically. To perform this analysis, we assume the \revision{sub-horizon approximation (SHA) and the quasi-static approximation (QSA)} and impose the \revision{longitudinal} gauge to single out scalar perturbations.

As a result we are able to obtain the values of the couplings in the QG action where the evolution of perturbations and the \revision{effective gravitational coupling} are free of instabilities, and where the cosmological evolution of such perturbations is well behaved. In such regimes, matter perturbations possess bounded solutions and follow standard solutions of second-order ordinary differential equations, and in the limiting super-horizon case, we re-obtain the GR behavior of perturbations. Outside these regions, however, reproducing standard cosmological evolution of perturbations is unfeasible due to the poles observed in the gravitational coupling of the theory.

We organize this work as follows: In section \ref{sec:section_2} we provide a brief review of cosmology in QG. In section \ref{sec:section_3} we derive the perturbed field equations for the FLRW metric and enforce the SHA and QSA to obtain the evolution equations for the perturbations. We then analyze the perturbations numerically and analytically, \revision{including their poles and limiting regimes}. In section \ref{sec:section_4} we discuss our results and point to future work.

Unless noted, geometric units are used, such that $c=\kappa=1$\revision{.}
\section{Cosmology in Quadratic Gravity}
\label{sec:section_2}
The gravitational action in QG, also called Stelle's action, can be written as

\begin{equation}
    \revision{\begin{aligned}
        S&=S_{\rm QG}+S_m,\\
        S_{\rm QG}&=\frac{1}{2\kappa}\int d^4x\sqrt{-g}\,
        \left[R + \alpha R^2 + \beta W^2\right],
    \end{aligned}}
    \label{eq:quadratic_action}
\end{equation}

where $R$ is the Ricci scalar and $W^2 = W_{\mu\alpha\nu\beta}W^{\mu\alpha\nu\beta}$ is the contraction of the Weyl tensor. \revision{The couplings have dimensions $[\alpha]=[\beta]=L^2$. We use $\kappa=1$ in the field equations below and restore $\kappa=8\pi G$ only when defining $G_{\rm eff}$.}
Such \revision{action} includes all possible quadratic terms \revision{in} curvature scalars \cite{Salvio:2018crh}, excluding topological terms. If the dynamical variable is given by the metric $g_{\mu\nu}$, the theory has fourth-order equations of motion, and Ostrogradsky's theorem implies the necessary existence of ghost modes of propagation \cite{Woodard:2015zca}. The existence of such ghosts, however, does not necessarily imply that the theory is globally unstable, as regions in the parameter space of the coupling constants $\alpha$ and $\beta$ allow local stability, and gauge fixing can constrain spurious degrees of freedom \cite{Salvio:2018crh,Alves:2025zkk}. 

Including a general stress energy tensor associated with minimally interacting matter action $S_m$, given by

\begin{equation}
    T^{\mu\nu} =  -\frac{2}{\sqrt{-g}}
\frac{\delta S_m}{\delta g^{\mu\nu}},
\label{eq:matter_stress_energy}
\end{equation}
\revision{Variation} of \eqref{eq:quadratic_action} plus the matter action \revision{with respect to} the metric $g_{\mu\nu}$ gives the general equation of motion for QG

\begin{align}
    G_{\mu\nu} &+ 2\alpha R\, R_{\mu\nu}
    -\frac{1}{2}\alpha\,R^2g_{\mu\nu} +2\alpha\,\Box R\,g_{\mu\nu}
    - 2\alpha\,\nabla_\mu\nabla_\nu R \notag\\
     &- \frac{1}{2}\beta\,W^2\,g_{\mu\nu}
     -\beta\,R_{\nu}^{\alpha\beta\gamma}\,W_{\mu\alpha\beta\gamma}
     - \beta\,R_{\mu}^{\alpha\beta\gamma}\,W_{\nu\alpha\beta\gamma}\notag\\
     &+ 4\beta\,W_{\mu}^{\alpha\beta\gamma}W_{\nu\alpha\beta\gamma}
     + 4\beta\, R^{\alpha\beta}W_{\mu\alpha\nu\beta}\notag\\
     &+ 2\beta\, \nabla_\alpha \nabla_\beta\, W_{\mu}\,^{\alpha}\,_{\nu}\,^{\beta}
     + 2\beta\, \nabla_\beta \nabla_\alpha W_{\mu}\,^{\alpha}\,_{\nu}\,^{\beta} \notag\\
    &= T_{\mu\nu},
    \label{eq:eom_quadratic}
\end{align}

where we keep the explicit dependence on the couplings $\alpha$ and $\beta$. 

In the case where $\beta\rightarrow 0$, one obtains the modified Einstein \revision{field equations} for a theory of the type $f(R) = R + \alpha R^2$, also called the Starobinsky model, if one takes $1/\alpha$ as the energy scale where the modified-gravity effects start to dominate \cite{Ketov:2025nkr}. By a suitable Legendre transform, such theories can be recast as \revision{scalar--tensor} theories, where there is only an extra scalar degree of freedom $\phi = f'(R)$.

The general FLRW metric is given by

\begin{equation}
    \begin{aligned}
    ds^2 ={}& -dt^2 + a^2(t)\left[\frac{dr^2}{1-Kr^2}\right.\\
    &\left.{}+r^2\left(d\phi^2 + \sin^2(\phi)d\theta^2 \right) \right],
    \end{aligned}
    \label{eq:FLRW_metric}
\end{equation}

where $k$ is the Gaussian curvature of the maximally symmetric 3-space. Such metric is conformally flat \cite{Islam:1992nt}, such that we have a vanishing Weyl tensor $W_{\mu\alpha\nu\beta} = 0$ and the e.o.m. is given by

\begin{align}
    G_{\mu\nu} &+2\alpha R\, R_{\mu\nu}
    -\frac{1}{2}\alpha \,R^2g_{\mu\nu}
    +2\alpha\,\Box R \,g_{\mu\nu}\notag\\
    &\revision{-2\alpha\nabla_\mu\nabla_\nu R} = T_{\mu\nu}.
    \label{eq:starobinsky_eom}
\end{align}

Solutions to equation ~\eqref{eq:starobinsky_eom} are known to be stable and possess no ghosts even in the presence of matter, since the action reduces to the $R+\alpha R^2$ form, which avoids Dolgov--Kawasaki and matter instabilities for $\alpha>0$ \cite{Sotiriou:2008rp,DeFelice:2010aj}. An immediate consequence of this is that cosmological solutions of \eqref{eq:starobinsky_eom} are stable at the background level in \revision{quadratic gravity}, such that instabilities may only arise from the extra massive spin-2 degree of freedom or from the dynamics of the perturbative solution. In particular, cosmological solutions for a perfect-fluid stress-energy tensor given by

\begin{equation}
    T_{\mu\nu} = \left(\rho + p\right)u_\mu u_\nu - pg_{\mu\nu}\,,
    \label{eq:stress_energy_perfect_fluid}
\end{equation}

where $u^\mu$, $\rho$ and $p$ are the 4-velocity, density and pressure of the energy component, respectively, have been widely explored for different components and equations of state, and an overview can be found in \cite{Capozziello:2018ddp}. 

To explore the cosmological evolution beyond the background cosmology, one needs to perturb the metric \eqref{eq:FLRW_metric} and the stress-energy tensor \eqref{eq:stress_energy_perfect_fluid} in order to derive the evolution of the energy content in the theory. Since perturbations of \eqref{eq:perturbed_flrw} do not lead to a conformally flat metric, the extra degrees of freedom associated with the quadratic equations of motion affect the evolution of such perturbations. We shall deal with such problem in the next section.

\section{Cosmological Perturbations}
\label{sec:section_3}
\subsection{Perturbative assumptions}

We now solve equation \eqref{eq:eom_quadratic} for linear perturbations of the FLRW metric \eqref{eq:FLRW_metric}, focusing on the scalar modes. One can check the full Scalar-Vector-Tensor decomposition of the perturbed equations of motion in \cite{Barrow:2007pm}. A dynamical systems analysis for the perturbed equations can be found in \cite{Barrow:2006xb} and in \cite{Asorey:2024oxw} for nonsingular solutions.

We assume the \revision{longitudinal gauge} in this work, such that the linear perturbations of the flat $K=0$ metric \eqref{eq:FLRW_metric} can be written in terms of the Bardeen potentials $\Phi$ and $\Psi$, which act similarly to the Newtonian potentials in the weak-field limit. In \cite{Salvio:2018crh} one can see that the two potentials are sufficient to describe the extra scalar degree of freedom of the theory, although they are related by a fourth-order equation due to the quadratic action. We follow a treatment of linear perturbations and approximations similar to the one found in \cite{Tsujikawa:2007gd}, where the stability of matter perturbations \revision{is analyzed under the QSA and SHA} for general scalar--tensor theories.

The full derivation of the equations has been validated
using \texttt{Mathematica} and notebooks with the step by step derivation using the \texttt{xAct} and \texttt{xPand} packages can be found in the github project \footnote{ \url{https://github.com/pedvoca/quadratic_cosmology} }.

The perturbed flat FLRW metric in the longitudinal gauge can be written as

\begin{align}
    ds^2 = -(1+2\Psi)dt^2 + a^2(t)\left(1-2\Phi \right)dx_i dx^i,
    \label{eq:perturbed_flrw}
\end{align}

and we also assume a linear, anisotropic perturbation of the stress-energy tensor \eqref{eq:stress_energy_perfect_fluid} for dust $p=0$, which is given by

\begin{align}
    &T_{\mu\nu} = \revision{\bar{\rho}_m} u_\mu u_\nu +\delta T_{\mu\nu}\,,\\
    &\delta T_{00} = \revision{\bar{\rho}_m\delta}\,,\quad \delta T_{0i}=\revision{\bar{\rho}_m v_i} \,,\quad\delta T_{ij} = 0,
    \label{eq:perturb_stress_energy}
\end{align}

where \revision{$\bar{\rho}_m$ is the background dust density and $\delta\equiv\delta\rho_m/\bar{\rho}_m$ is the dimensionless matter-density contrast.}

In order to control the propagation modes, we focus on modes well inside the Hubble horizon,
\begin{equation}
 \frac{k^2}{a^2H^2}\gg1,
 \label{eq:late_subhorizon_condition}
\end{equation}
the so-called \revision{sub-horizon approximation (SHA)}. Such an approximation allows us to \revision{neglect} terms proportional to $H$ in the Bardeen \revision{potentials} in relation to the \revision{dominant} modes $k/a$, since we focus on modes propagating \revision{within} a Hubble radius.

We furthermore make the assumption of the \revision{quasi-static approximation (QSA)}, which assumes that 

\begin{equation}
	\nabla_i X \gg \nabla_0 X,
\label{eq:quasi_static_approximation}
\end{equation}
for a perturbation $X$. Physically, this means that the time variation of perturbations is small compared to their local spatial variations. This hypothesis is well motivated in matter dominated Universes, as the Bardeen potentials are constant in time during matter domination for anisotropic perturbations \cite{durrer_2020}.

 An extensive analysis of both the SHA and QSA, can be found in \cite{Orjuela-Quintana:2023zjm} in the case of $f(R)$ theories. As shown in the previous section, the quadratic terms become relevant only at the perturbative level, such that they are subdominant in relation to the $R^2$ term at all orders, and their propagation modes are under control following the dominant terms.


\subsection{Linear Perturbation Equations}
\label{sec:late_time_scalar_reduction}

By defining the left side of equation \eqref{eq:eom_quadratic} as being $\mathcal E_{\mu\nu}$, the $00$ and traceless $ij$ equations of motion, under the SHA  reduce to
\begin{align}
&\begin{gathered}
 4\alpha\Box^2(\Psi-2\Phi)
 +\frac{4}{3}\beta\Box^2(\Psi+\Phi)
 +2\Box\Phi
 =\delta\rho_m,\\[-2pt]
 \left(\mathcal E_{00}=T_{00}\right)
\end{gathered}
\label{eq:late_qg_00_box}\\
&\begin{gathered}
 -\frac14\nabla_{\langle i}\nabla_{j\rangle}(\Psi-\Phi)
 +\alpha\nabla_{\langle i}\nabla_{j\rangle}\Box(\Psi-2\Phi)\\[-2pt]
 -\frac{\beta}{6}\nabla_{\langle i}\nabla_{j\rangle}
     \Box(\Psi+\Phi) =0,\\[-2pt]
 \left(\mathcal E_{\langle ij\rangle}=T_{\langle ij\rangle}\right).
\end{gathered}
\label{eq:late_qg_ij_box}
\end{align}

Here $_{\langle\rangle}$ is used to indicate we take the traceless component of the corresponding indices.  For the scalar dust perturbation $\delta $ and the velocity 3-vector  $v^i=\nabla^i v$ defined in \eqref{eq:perturb_stress_energy}, we obtain the continuity and Euler equations for the matter perturbations following the conservation of the stress-energy tensor,

\begin{align}
&\dot{\delta\rho}_m+3H\delta\rho_m
 =\bar\rho_m\left(3\dot\Phi+D_i v^i\right),\qquad &&\left(\nabla^{0}T_{00} = 0\right)
 \label{eq:late_dust_continuity_cosmic}\\
&\dot v+Hv=\frac{\Psi}{a}. \qquad&& \left(\nabla^{i}T_{i0} = 0\right)
\label{eq:late_dust_euler_cosmic}
\end{align}
Equations \eqref{eq:late_qg_00_box}--\eqref{eq:late_dust_euler_cosmic} provide the full equations for the evolution of the scalar dust perturbation $\delta$, given suitable initial conditions. 

We transform to Fourier space in the spatial componentes, using the negative convention for the covariant Laplacian $\Delta \equiv \nabla^i\nabla_i$

\begin{equation}
	 \Delta X  \longleftrightarrow -\frac{k^2}{a^2} X,
	\label{eq:fourier_transform}
\end{equation}  

for a tensor field $X$. 
Given the above convention, we can combine equations \eqref{eq:late_dust_continuity_cosmic} and \eqref{eq:late_dust_euler_cosmic}, to obtain the second order equation for the evolution of the matter perturbations in Fourier space,

\begin{equation}
 \ddot\delta+2H\dot\delta+\frac{k^2}{a^2}\Psi
 =3\left(\ddot\Phi+2H\dot\Phi\right).
 \label{eq:late_growth_exact_scalar}
\end{equation}

By using the QSA, we neglect the time derivatives on the right side of \eqref{eq:late_growth_exact_scalar} to arrive at

\begin{equation}
 \ddot\delta+2H\dot\delta+\frac{k^2}{a^2}\Psi\simeq0.
 \label{eq:late_growth_before_qg}
\end{equation}

Equation \eqref{eq:late_growth_before_qg} gives the standard second order equation for the evolution of matter perturbations, although it still depends on the behavior of the Bardeen potential $\Psi$, such that one still needs an extra relation between $\Psi$ and $\delta$ to obtain the evolution of the matter perturbation. In order to obtain such relation, we go back to the Modified Einstein Equations \eqref{eq:late_qg_00_box}--\eqref{eq:late_qg_ij_box}.

By going to Fourier space, equations \eqref{eq:late_qg_00_box}--\eqref{eq:late_qg_ij_box} become

\begin{align}
 4\alpha\frac{k^4}{a^4}(\Psi-2\Phi)
 +\frac{4}{3}\beta\frac{k^4}{a^4}(\Psi+\Phi)
 -2\frac{k^2}{a^2}\Phi
 &=\delta\rho_m,
 \label{eq:late_qg_00_fourier}\\
 -\frac14(\Psi-\Phi)
 -\alpha\frac{k^2}{a^2}(\Psi-2\Phi)
 +\frac{\beta}{6}\frac{k^2}{a^2}(\Psi+\Phi)
 &=0.
 \label{eq:late_qg_ij_fourier}
\end{align}

We factor out the \revision{nonzero} operator
$\nabla_{\langle i}\nabla_{j\rangle}$ in \eqref{eq:late_qg_ij_fourier}, which in Fourier space gives terms $\propto k^2/a^2$ . Such factoring is suitable given that $k/a>0$ for the propagation modes under analysis and, in fact, physical modes.

Multiplying \eqref{eq:late_qg_ij_fourier} by $-4$ gives
\begin{equation}
\left(1+4\alpha\frac{k^2}{a^2}
       -\frac{2}{3}\beta\frac{k^2}{a^2}\right)\Psi
 -\left(1+8\alpha\frac{k^2}{a^2}
       +\frac{2}{3}\beta\frac{k^2}{a^2}\right)\Phi=0.
 \label{eq:late_slip_expanded}
\end{equation}

Solving for $\Psi$ gives
\begin{equation}
 \Psi=
 \frac{1+8\alpha\frac{k^2}{a^2}
       +\frac{2}{3}\beta\frac{k^2}{a^2}}
      {1+4\alpha\frac{k^2}{a^2}
       -\frac{2}{3}\beta\frac{k^2}{a^2}}\,
 \Phi,
 \label{eq:late_psi_of_phi}
\end{equation}
which naturally motivates us to define the gravitational slip function
\begin{equation}
 Q(k,a)\equiv\frac{\Phi}{\Psi}
 =
 \frac{1+4\alpha\frac{k^2}{a^2}
       -\frac{2}{3}\beta\frac{k^2}{a^2}}
      {1+8\alpha\frac{k^2}{a^2}
       +\frac{2}{3}\beta\frac{k^2}{a^2}},
 \label{eq:slip_function}
\end{equation}
which matches the usual definition of the slip function in Modified Gravity \cite{Clifton:2011jh}. Using the slip function in \eqref{eq:late_qg_00_fourier}, we obtain the equation


\begin{equation}
 \delta\rho_m
 =\frac{k^2}{a^2}\Psi
 \left[
 4\alpha\frac{k^2}{a^2}(1-2Q)
 +\frac{4}{3}\beta\frac{k^2}{a^2}(1+Q)
 -2Q
 \right],
 \label{eq:delta_psi_relation}
\end{equation}

which gives us an algebraic relation (in Fourier space) between the Bardeen potential $\Psi$ and the matter density perturbation $\delta\rho$. From this relation we are finally able to analyze the behavior of the perturbation under the SHA and QSA assumptions.

\subsection{Evolution of Matter Perturbations}

Equation \eqref{eq:delta_psi_relation} is a modified Poisson equation for the matter density perturbation in Fourier space, which motivates us to define the functional form of the modification multiplying the gravitational potential. We define
\begin{equation}
 f_Q(k,a)\equiv
 \left[\begin{aligned}
 &4\alpha\frac{k^2}{a^2}\bigl(1-2Q(k,a)\bigr)\\
 &+\frac{4}{3}\beta\frac{k^2}{a^2}\bigl(1+Q(k,a)\bigr)
 -2Q(k,a)
 \end{aligned}\right]^{-1},
 \label{eq:f_q_definition}
\end{equation}
which gives \revision{us} the relation 

\begin{equation}
 \frac{k^2}{a^2}\Psi=f_Q(k,a)\,\delta\rho_m
 =f_Q(k,a)\,\bar\rho_m\delta.
 \label{eq:poisson_f_q}
\end{equation}

Relation \eqref{eq:poisson_f_q} provides the effective gravitational coupling function for cosmological perturbations in QG. The function $f_Q(k,a)$ contains all of the extra quartic-order terms in $k/a$, and in the GR limit $\alpha,\beta \rightarrow 0$, we have $f_Q(k,a)=-1/2$, returning to the Poisson equation for a matter overdensity in Newtonian gravity.

By substitution in \eqref{eq:late_growth_before_qg} we can finally write equation \eqref{eq:delta_psi_relation} as a closed differential equation for the matter perturbation:

\begin{equation}
 \ddot\delta+2H\dot\delta
 +f_Q(k,a)\,\bar\rho_m\delta\simeq0.
 \label{eq:late_growth_fQ}
\end{equation}

Equation \eqref{eq:late_growth_fQ} gives the evolution of the perturbation $\delta$ in QG for universes with pressureless energy content $w=0$. By restoring the dimensionful quantity $\kappa = 8\pi G$, we can explicitly write the evolution equation with the effective gravitational coupling:

\begin{equation}
\ddot{\delta} + 2H\dot{\delta}
- 4\pi G_{\rm eff}\,\bar{\rho}_m \delta
\simeq 0,
\label{eq:late_Geff}
\end{equation}

where the effective gravitational \revision{coupling} is given by
\begin{equation}
G_{\rm eff}(k,a)
= -2 f_Q(k,a)\,G,
\label{eq:geff_coupling}
\end{equation}
which can be directly compared to the effective gravitational coupling in scalar--tensor theories as found in \cite{Tsujikawa:2007gd}. 

\begin{figure*}[t]
	\centering
    \includegraphics[width=0.86\textwidth]{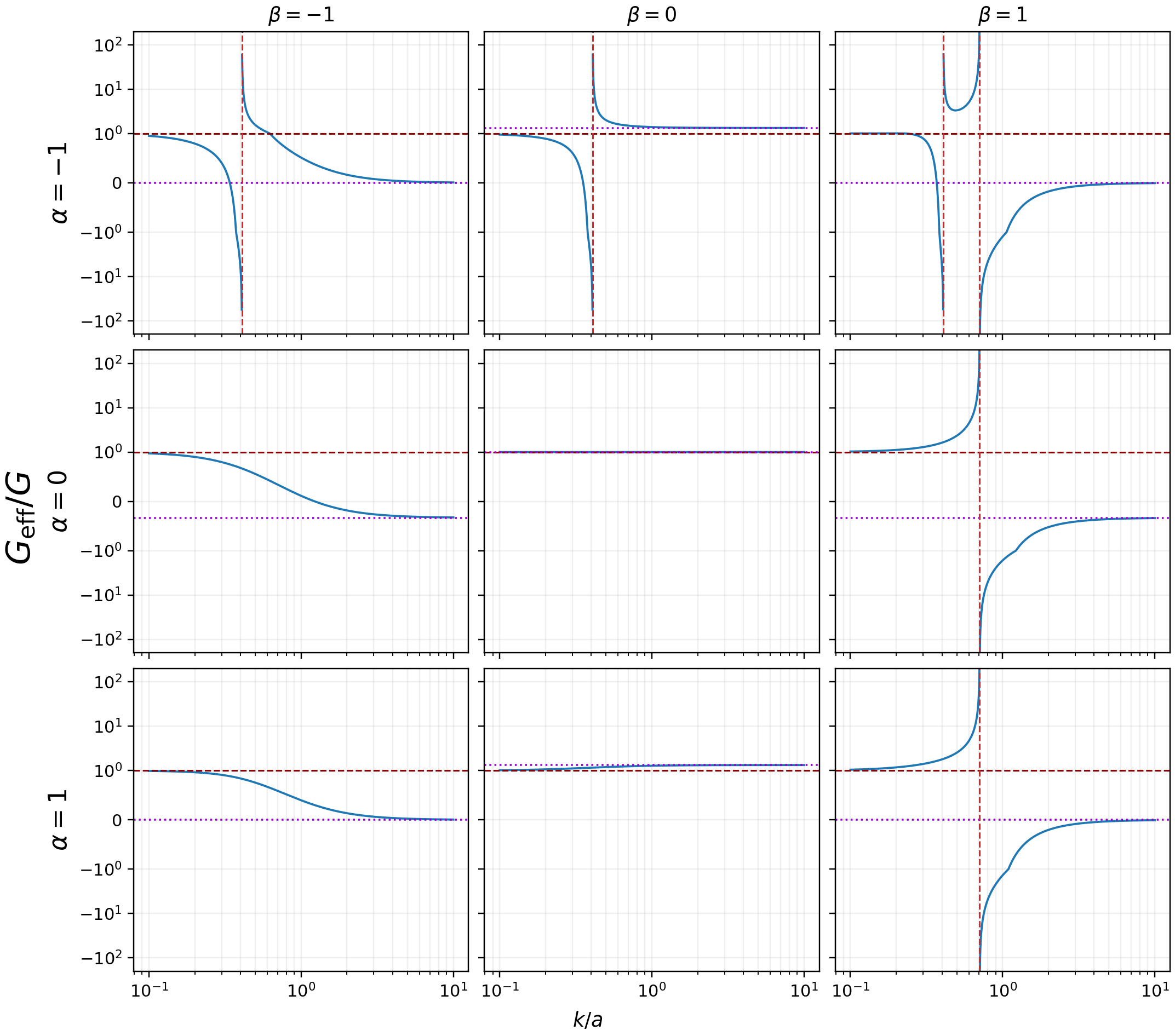}
	\caption{Scale dependence of the effective gravitational coupling
		$G_{\rm eff}/G$ as a function of the physical wavenumber $k/a$. Rows correspond
		to $\alpha=(-1,0,1)$ and columns to $\beta=(-1,0,1)$. Solid blue curves show
		$G_{\rm eff}/G$; red vertical dashed lines denote poles on the real $k/a$ axis.
		The dark-red dashed horizontal line is the universal formal infrared limit,
		$G_{\rm eff}/G\to1$ as $k/a\to0$, equivalently $f_Q\to-1/2$. Violet dotted
		lines denote the ultraviolet limits: $G_{\rm eff}/G\to0$ for
		$\alpha\beta\neq0$, $4/3$ for $\beta=0$, $-1/3$ for $\alpha=0$, and $1$ for
		$\alpha=\beta=0$. The vertical axis uses a symmetric logarithmic scale.}
    \label{fig:G_eff_parameters}
\end{figure*}

In figure \ref{fig:G_eff_parameters} we show the behavior of $G_{\rm eff}(k/a)$ for different values of the parameters $\alpha$ and $\beta$. \revision{The poles of $G_{\rm eff}$ are the zeros of its denominator, equivalently of the inverse bracket defining $f_Q$ in \eqref{eq:f_q_definition}; they are not zeros of $G_{\rm eff}$. Since the pole structure depends on the signs of the couplings, we use the representative values $\{-1,0,1\}$. The resulting denominator is an even polynomial in the complex variable $k/a$ of degree at most four, with roots that are either real or purely imaginary. Restricting to positive real roots, which correspond to the physical modes $k/a>0$, gives no poles when $\alpha\geq0$ and $\beta\leq0$, two poles when $\alpha<0$ and $\beta>0$, and one pole otherwise.} In the $\beta=0$ case, $G_{\rm eff}$ behaves exactly as in $f(R)$ gravity, interpolating between the GR and Brans--Dicke behavior shown in \cite{Tsujikawa:2007gd}, while $G_{\rm eff}$ shows \revision{a sign change} near poles in the $\alpha<0$ case and whenever $|\beta|>1$.

It should be noted that the poles are always found in regions where $k/a\sim 1$, such that they could point to a breakdown of the SHA and the need for a fully nonlinear analysis of the equation, instead of a pathological behavior of the theory. In any case, it is clear that linear perturbations are not globally stable for general values of the parameters, with the \revision{$\alpha\geq0$ and $\beta\leq0$} regions avoiding poles in $G_{\rm eff}$.
 
\begin{figure*}[t]
	\centering
    	\includegraphics[width=0.92\textwidth]{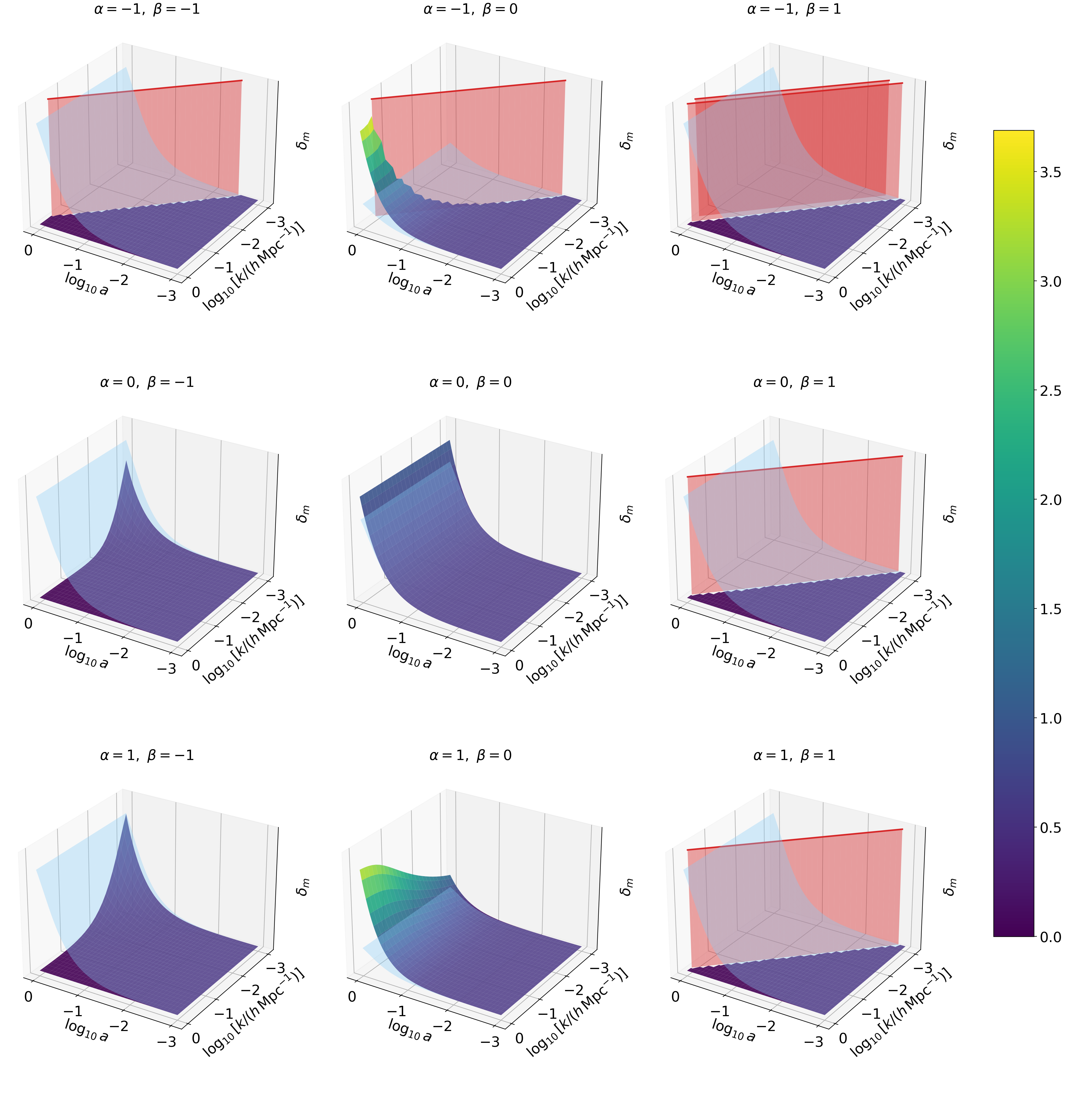}
	\caption{Three-dimensional numerical solutions for the growing matter-density perturbation $\delta_m(k,a)$ in quadratic gravity, shown for $\alpha,\beta\in\{-1,0,1\}$. The horizontal axes are $\log_{10}a$ and $\log_{10}[k/(h\,{\rm Mpc}^{-1})]$, \revision{where $h\equiv H_0/(100\,{\rm km}\,{\rm s}^{-1}\,{\rm Mpc}^{-1})$ is dimensionless}, while the vertical height and the shared colour bar encode $\delta_m$; numerical tick labels on the vertical axes are omitted. The quadratic-gravity solutions assume a matter-dominated background, $\Omega_{m0}=1$ and $\Omega_{\Lambda0}=0$, with initial conditions $\delta_m(a_{\rm ini})=a_{\rm ini}$ and $\delta_{m,a}(a_{\rm ini})=1$. The translucent light-blue surface is the scale-independent $\Lambda$CDM growing solution, evaluated with $\alpha=\beta=0$, $\Omega_{m0}=0.3$, and $\Omega_{\Lambda0}=0.7$, using the same initial normalization. Translucent red sheets mark the positive-real poles of the effective gravitational coupling, \revision{defined by the vanishing denominator of $G_{\rm eff}$}. For each mode, the numerical integration is terminated at the first pole encountered; no continuation through the singularity is performed.}
		 \label{fig:delta_m_solutions}
\end{figure*}

 \revision{Away from poles of $G_{\rm eff}$, equation \eqref{eq:late_Geff} has continuous coefficients for a continuous $H$ and therefore admits a unique local solution once initial data are specified. This local existence statement does not establish global or nonlinear stability.} In particular, we solve \eqref{eq:late_Geff} numerically away from the poles for explicit couplings and scale $(\alpha,\beta,k/a)$. In figure \ref{fig:delta_m_solutions} we show the \revision{three-dimensional} surfaces defined by the growing solutions $\delta(k,a)$ of equation \eqref{eq:late_Geff}, for representative values of the couplings $(\alpha, \beta)$, as well as the surfaces defined by the poles of each growing solution. We truncate the solutions once the numerical solver encounters a pole. For comparison, we plot the evolution of matter-density perturbations in a \revision{$\Lambda$CDM} universe with $\Omega_m = .3$, $\Omega_\Lambda= 1-\Omega_m$ in light blue.

As expected, the pole behavior follows what was found in \eqref{eq:late_Geff}, and we focus on the behavior of the growing solution $\delta_m$. Comparing with the $\Lambda$CDM growth found in light blue, we see that for $\beta>0$, the solutions necessarily go through a pole from the matter-dominated epoch to $a=1$, such that these solutions cannot reproduce the observed cosmological behavior. For $\beta\leq0$, the \revision{stable} region $\alpha\geq0$ shows an evolution of the matter perturbation capable of reproducing $\Lambda$CDM behavior, although in the case of negative $\beta$, such evolution is found at high $k/a$, where the theory approaches the scalar--tensor-theory behavior. The $\beta=0$ region returns to the well-understood $f(R)$-gravity behavior, capable of reproducing stable late-time cosmological evolution when $\alpha\geq0$, as discussed in the previous section.

\subsection{Analytical Limits}

In this subsection we focus on obtaining analytic results in the deep \revision{sub-horizon} limit $k/a\rightarrow{}\infty$ and the non-rigorous super-horizon limit $k/a\rightarrow{}0$, which explicitly violates the SHA; nonetheless, its analysis is valuable as it provides a test for the limit of such approximation and \revision{supports} the numerical results obtained in the previous subsection.

\subsubsection{Deep \revision{Sub-Horizon} Limit}

In equation \eqref{eq:late_Geff}  when the modes are deep within the horizon, we have

\begin{equation}
 \begin{aligned}
 \frac{k}{a}\longrightarrow \infty:\qquad
 Q&\longrightarrow \frac{6\alpha - \beta}{12\alpha + \beta},\\
 f_Q&\longrightarrow0,\qquad
 G_{\rm eff}\longrightarrow0,
 \end{aligned}
 \label{eq:late_fQ_uv}
\end{equation}
as long as the couplings satisfy $\alpha\beta\neq 0$. 

In such case, we find that the effective gravitational coupling vanishes, and the evolution of the perturbations is determined solely by the evolution of the Hubble parameter $H$. Equation  \eqref{eq:late_growth_fQ} becomes

\begin{equation}
 \ddot\delta+2H\dot\delta=0,
 \label{eq:late_growth_uv}
\end{equation}

which has the \revision{leading behavior}

\begin{equation}
\revision{\delta(t)\sim\int_{t_0}^{t}dt'\,
\exp\!\left[-2\int_{t_0}^{t'}dt''\,H(t'')\right].}
\label{eq:subhorizon_perturbation}
\end{equation}

\revision{Thus, the leading ultraviolet behavior is determined entirely by the background expansion rate $H(t)$, without specifying its functional form.} We note how such limit differs from standard $f(R)$ theory, where in the deep sub-horizon limit the perturbation dynamics are given by a Brans--Dicke theory with parameter $\omega = 0$ \cite{Tsujikawa:2007gd}. In field-theoretic language, in FLRW metrics the ultraviolet behavior of the theory fundamentally differs from theories of the type $R+\alpha R^2$ due to the extra massive spin-2 degree of freedom. 

\subsubsection{Super-Horizon \revision{Limit}}

In the (non-rigorous) super-horizon limit we obtain, from \eqref{eq:late_Geff},

\begin{equation}
 \frac{k}{a}\longrightarrow0:
 \qquad
 Q\longrightarrow1,
 \qquad
 f_Q\longrightarrow-\frac12,
 \qquad
 G_{\rm eff}\longrightarrow G
 \label{eq:late_fQ_ir},
\end{equation}

such that equation \eqref{eq:late_growth_fQ} becomes

\begin{equation}
\ddot\delta + 2H\dot\delta +  4\pi G\bar\rho\,\delta = 0,
\label{eq:superhorizon_perturbation}
\end{equation}

and the perturbations behave exactly as in GR. The evolution of such perturbations is thoroughly examined in standard cosmology textbooks and can be found, e.g., in \cite{durrer_2020}.

Although the limit \eqref{eq:late_fQ_ir} violates the SHA, it shows that one can safely truncate equation \eqref{eq:late_Geff} to leading order and obtain stable behavior from the perturbations. Such result points to the \revision{adequacy} and stability of the perturbative assumption under the SHA for QG. We also find that the infrared behavior of the theory returns to the GR regime as expected, which points to the possibility of finding stable perturbations beyond the SHA given that we re-obtain the correct infrared behavior of the theory. This is, as previously mentioned, dependent on the value of the parameters $(\alpha,\beta)$.

\section{Discussion and Conclusion}
\label{sec:section_4}
In this paper we have analyzed the stability and evolution of the effective gravitational \revision{coupling $G_{\rm eff}$} and linear matter-density perturbations $\delta_m$ in QG. We \revision{use} both the \revision{SHA and QSA}, \revision{derive} the QG field equations to first order in the FLRW metric, which we validated using \revision{\texttt{Mathematica}}, and obtain the second-order differential equation describing the evolution of cosmological matter perturbations in the theory. We obtain a closed form \revision{for} the effective gravitational \revision{coupling $G_{\rm eff}(k,a)$} and identify its poles. An analysis of these poles shows that QG can reproduce the observed cosmological evolution of linear matter-density perturbations for parameters $\alpha\geq 0$ and $\beta\leq0$ and, in particular, for small values of $k/a$. Such behavior is close to the already well studied and understood $f(R)$ regime. Poles are found close to $k/a\sim1$, which points to the possibility \revision{of a breakdown} of the linear, quasi-static, or sub-horizon approximations, instead of an intrinsic instability of the theory. The full linear analysis of such equations, without assumptions on the QSA or SHA, is left for future work.

Our analytical results show that to first order in the FLRW metric the theory recovers the infrared behavior of GR as expected, and in the ultraviolet the gravitational coupling $G_{\rm eff}$ vanishes, such that perturbations \revision{have the leading behavior given by \eqref{eq:subhorizon_perturbation}, determined by $H(t)$ rather than free oscillations.} This significantly departs from the behavior found in $R^2$ and general $f(R)$ theories.
In conclusion, our analysis shows an intrinsic difficulty of QG in reproducing late-time cosmological behavior as observed due to the introduction of new poles in the effective gravitational \revision{coupling}, such that the extra degrees of freedom hinder the ability of the theory to properly describe linear perturbations in an expanding Universe.
\section*{Disclaimer}
\revision{Artificial-intelligence tools were used to improve the text and to assist with validation of calculations. The calculations and numerical results were independently checked using Python and Mathematica.}
\bibliography{main}
\end{document}